\newcommand{\vecr}{\mbox{\boldmath$r$}}
\newcommand{\vecv}{\mbox{\boldmath$v$}}
\newcommand{\vecp}{\mbox{\boldmath$p$}}
\newcommand{\vece}{\mbox{\boldmath$e$}}
\newcommand{\veca}{\mbox{\boldmath$a$}}
\newcommand{\vecu}{\mbox{\boldmath$u$}}
\newcommand{\vecg}{\mbox{\boldmath$g$}}

\newcommand{\vecK}{\mbox{\boldmath$K$}}
\newcommand{\vecF}{\mbox{\boldmath$F$}}
\newcommand{\vecnl}{\mbox{\boldmath$0$}}

\newcommand{\vecomega}{\mbox{\boldmath$\omega$}}
\newcommand{\dfd}{{\rm d}}

\documentclass{article}

\begin{document}

\title{Mechanics, cosmology and Mach's principle}
\author{Hanno Ess\'en
\\KTH Mechanics  \\ SE-100 44 Stockholm, Sweden\\
e-mail: hanno@mech.kth.se}
\date{November 2012}
\maketitle

\begin{abstract}
It is pointed out that recent cosmological findings seem to support the view that the mass/energy distribution of the universe defines the Newtonian inertial frames as originally suggested by Mach. The background concepts of inertial frame, Newton's second law, and fictitious forces are clarified. A precise definition of Mach's principle is suggested. Then an approximation to general relativity discovered by Einstein, Infeld, and Hoffmann is used and it is found that this precise formulation of Mach's principle is realized provided the mass/energy density of the universe has a specific value. This value turns out to be twice the critical density. The implications of this approximate result is put into context.
\end{abstract}

\section{Introduction}
In 2011 the final report from the Gravity probe B experiment was published by Everitt et al.\ \cite{everitt&al}. This satellite experiment verified for the first time the frame dragging prediction of general relativity thereby corroborating one aspect of Mach's principle. This principle states that the inertial frames of classical mechanics are defined as being those that do not accelerate with respect to the average mass/energy distribution of the universe. Since the early work on Mach's principle by Bondi \cite{BKbondi,bondi}, Sciama \cite{BKsciama, sciama, sciama&al}, Dicke \cite{dicke}, Lynden-Bell \cite{lyndenbell2}, and others, cosmology has progressed considerably. The cosmic microwave background radiation has been discovered and studied in detail. Dark matter has been found to dominate over normal matter in the universe. The recent discovery of the acceleration of the Hubble expansion by the groups of Perlmutter and of Schmidt and Riess (Nobel prize 2011) has given new observational information on the mass/energy distribution of the universe. In particular it is now clear that matter, dark matter, and dark energy together represent a mass/energy density close to the critical density of cosmology \cite{BKcheng_tp,jordan}. It seems appropriate to reassess the status of Mach's principle in view of these empirical advances of recent decades.

After a brief survey of the most relevant literature I first discuss Newton's second law and its interpretation. The concept of inertial frame as well as the related ideas about real and fictitious forces are elucidated. This part can be understood by undergraduates who have studied vector mechanics and tries to give a deeper motivation for the various terms that appear in this law. Then I demonstrate how this law arises by approximation from general relativistic equations of motion, and how this seems to support Mach's principle as a consequence. This part relies on the Einstein-Infeld-Hoffmann Lagrangian formulation and requires that the students have been exposed to analytical mechanics.

Mach's principle has been subject to much discussion and speculation in the literature. Some more recent studies can be found in the volume edited by Barbour and Pfister \cite{BKbarbour&pfister}, and in Mashhoon et al.\ \cite{mashhoon&al} where the gravitomagnetic analogy is studied in detail. Many points of view are of a philosophical and metaphysical nature, but here I will concentrate on empirical aspects. There are quite a few accounts also in the pedagogical literature, {\it e.g.}\ Zylbersztanj \cite{zylbersztanj}. Interesting discussions can be found in the textbooks by Berry \cite{BKberry}, Ciufolini and Wheeler \cite{BKciufolini&wheeler}, Peacock \cite{BKpeacock}, and, most recently, by Cheng \cite{BKcheng_tp}. Frame dragging and its relation to Mach's principle and general relativity has been discussed by Gr{\o}n \cite{gron,hartman}, Gr{\o}n and Eriksen \cite{gron&eriksen}, Harris \cite{harris}, Holstein \cite{holstein}, Hughes \cite{hughes}, Lynden-Bell et al. \cite{lyndenbell&bicak&katz}, Mart\'{i}n et al.\ \cite{martin&al}, Nightingale \cite{nightingale,nightingale&ray},  and by Vet\H{o} \cite{veto1,veto2}. It has been pointed out that general relativity does not, in itself, imply Mach's principle, the counter example being G\"{o}del's solution \cite{rindler} to Einstein's equations. In this solution, however, time travel is possible. There have been speculations that banning time travel will restrict us to solutions that obey Mach's principle \cite{laurent}, but these matters are still far from clear.

\section{Inertial frames, real and fictitious forces}
Newton's second law for a particle is nowadays normally written in the form
\begin{equation}\label{eq.newton2.inert}
m \veca = \vecF.
\end{equation}
Here $m$ is (inertial) mass which is determined by means of a common balance and a reference mass (unit of mass). The acceleration $\veca$ is a purely kinematic quantity which is determined by recording the positions with respect to time relative to a chosen {\it reference frame}. This requires the choice of origin, axes, measuring rods and clocks (reliable periodic phenomena). The force $\vecF$ then turns out to be a quantity which is the cause of the acceleration. All known forces decay at least as the inverse distance squared, so that forces have local origin, {\it i.e.}\ they arise from {\it local sources} near the moving particle and they obey Newton's third law of action and reaction. The success of Newtonian mechanics comes largely from the fact that there is a limited catalogue of forces. We have very accurate mathematical models for electromagnetic and gravitational forces. All other forces of macroscopic importance are material contact forces (normal force, dry friction, pressure, viscosity, drag, force from elastic deformation, pull from a string, {\it etc}) for which there often are reasonably accurate mathematical models, albeit of limited range of validity.

In the paragraph above I have emphasized the words reference frame and local source. In order for the above theory to work the reference frame must be so called {\it inertial}, or non-accelerated. We can use accelerated frames, however, if we know their translational acceleration $\veca_{\rm f}$ and their angular velocity $\vecomega$ about an axis through the origin. The appropriate equation of motion is then
\begin{equation}\label{eq.newton2.non.inert}
m \veca^* = \vecF + \vecF^* ,
\end{equation}
where,
\begin{equation}\label{eq.fict.force}
\vecF^* = -m [\veca_{\rm f} + \vecomega\times(\vecomega\times\vecr^*)+2\,\vecomega\times\vecv^* + \dot{\vecomega}\times\vecr^*].
\end{equation}
Here $\vecr^* , \vecv^*,$ and $\veca^*$ are position, velocity, and acceleration with respect to the accelerating system, while $\dot{\vecomega}$ is the time derivative of the angular velocity vector. The new force $\vecF^*$ on the right hand side of (\ref{eq.newton2.non.inert}) is the vector sum of the so called fictitious forces (\ref{eq.fict.force}). These forces do not have local sources and Newton's third law does not apply to them. The crucial question is now: acceleration relative to what? Which are the inertial frames? How are they found?

When it comes to rotation this question is fairly easy to answer. In a non-rotating reference frame the fixed stars (or even better, the distant galaxies) have fixed directions. Since Earth rotates with respect to such a frame the fictitious forces containing the angular velocity $\vecomega$ are necessary to get the correct motion of {\it e.g.}\ a Foucault pendulum. Also the fact that rotating astronomical bodies are flattened is well accounted for by the fictitious centrifugal force $-m\,\vecomega\times(\vecomega\times\vecr^*)$. It is the tiny difference between the angular velocity of the Earth relative to the fixed stars and the angular velocity appearing in these equations that was discovered by Gravity Probe B \cite{everitt&al}. The inertial frames near the Earth rotate slightly because the Earth rotates, according to general relativity. Today there is no observational evidence that our universe is a rotating G{\"o}del universe \cite{rindler}.

The question is much more complicated when it comes to translational acceleration. When studying motion in a laboratory on Earth we are used to including the gravitational force $m \vecg$ arising from our planet. If we denote the non-gravitational force on our particle by $\vecK$ we then normally write down the equation of motion
\begin{equation}\label{eq.normal.eq.of.m}
m\veca = \vecK + m\vecg .
\end{equation}
A critical person may now note that the particle, in fact, must also be affected by the gravitational forces from the Sun, the Moon, the Galaxy, and so on. Let us denote the acceleration that these gravitational forces would impart to our particle by $\vecg_{\rm f}(\vecr)$ at position $\vecr$. To get an accurate equation of motion $m\vecg_{\rm f}$ must then be added on the right hand side of (\ref{eq.normal.eq.of.m}).

We now realize, however, that the reference frame at rest on the Earth is not really inertial (even neglecting rotation). Earth is freely falling in the external gravitational fields from bodies other than the Earth. This acceleration will vary with position, but not that much. Let us choose the value at the origin and put $\vecg_{\rm f}(\vecnl)=\veca_{\rm f}$. The true equation of motion for a particle on a ( non-rotating) freely falling Earth is then,
\begin{equation}\label{eq.newton2.non.inert.earth}
m \veca^* = \vecK + m \vecg + m \vecg_{\rm f}(\vecr) - m \veca_{\rm f} \approx  \vecK + m \vecg .
\end{equation}
The approximation can be made since $\vecg_{\rm f}(\vecr) \approx \veca_{\rm f}$ for $\vecr$-values of interest. The fact that there is exact equality only at a point gives rise to tidal effects. The Earth falls freely in the field of the Sun and the Moon but since the Earth is extended all points are not subject to the same acceleration $\veca_{\rm f} = \vecg_{\rm f}(\vecnl)$.

To summarize, when working in a lab on Earth I am in fact working in an accelerated reference frame that accelerates in such a way that gravitational forces $m \vecg_{\rm f}(\vecr)$ from other bodies than the Earth itself are transformed away by a fictitious force $\vecF^* = - m \veca_{\rm f} = -m \vecg_{\rm f}(\vecnl)$. The usual equation of motion (\ref{eq.normal.eq.of.m}) thus works quite well, but the $\veca$ occurring in it is in fact relative to an accelerated reference frame and is thus really an $\veca^*$.

The precise acceleration of the Earth relative to the universe as a whole is quite difficult to measure. The phenomenon of Doppler shift, however, makes it possible to find the velocity of the Earth with respect to the cosmic microwave background (CMB) quite accurately. Such measurements reveal that constant velocity with respect to the CMB seems to correspond to inertial frames. This possibility of identifying the rest frame of the CMB (the frame in which the radiation is as isotropic as possible) with the frame in which the mass/energy of the universe as a whole is at rest, constitutes an observational verification of Mach's principle (see Cheng \cite{BKcheng_tp}, Sec.\ 10.5.4).

\section{Newton's second law and Einstein's equations}
Mach's principle was one of the inspirations behind Einstein's work on general relativity but the precise connection is still not clear. Einstein discusses the connection in his book {\it The Meaning of Relativity} \cite{BKeinstein} and outlines how the cosmic mass/energy density influences the equations of motion of a particle (on pages 100-102). A detailed derivation of these equations is given by Harris \cite{harris}.

Let us make Mach's principle more precise. Assume that the equation of motion of a particle is of the form,
\begin{equation}\label{eq.mach.princip.precise}
m (\veca - \vecg_{\rm u}) = \vecF ,
\end{equation}
where $\vecg_{\rm u}$ is the acceleration of the universe as a whole. Then only the acceleration relative to the universe as a whole is what matters in the equation of motion. I now proceed to show that this, at least is a possibility.

I will approach this problem from the point of view of the Einstein-Infeld-Hoffmann (EIH) equations of motion \cite{EIH,eddington&clark,BKinfeld&plebanski}. Fock found the Lagrangian that yields these equations \cite{BKfock} and this Lagrangian is also derived and discussed in Landau and Lifshitz \cite{BKlandau2}, Brumberg \cite{brumberg}, and Kennedy \cite{kennedy}. The Lagrangian is given by
\begin{equation}\label{eq.eih.lagr}
L = L_0 + L_1 + L_2 ,
\end{equation}
where,
\begin{equation}\label{eq.eih.lagr0}
L_0 = T_0-V_0= \sum_a \frac{1}{2} m_a \vecv_a^2 + \frac{1}{2} \sum_a \sum_{b\neq a}  \frac{G m_a m_b}{r_{ab}} ,
\end{equation}
and
\begin{equation}\label{eq.eih.lagr1}
L_1 =
\frac{1}{4c^2} \sum_a \sum_{b\neq a}  \frac{G m_a m_b}{r_{ab}} \left[ 3(\vecv_a^2+\vecv_b^2)-7\vecv_a\cdot\vecv_b - \frac{(\vecv_a\cdot\vecr_{ab}) (\vecv_b\cdot\vecr_{ab})}{r_{ab}^2} \right].
\end{equation}
$L_2$ contains the first relativistic correction, $\sim (v/c)^2$, to the classical kinetic energy $T_0$, and higher order corrections to the gravitational interaction $\sim G^2$. These terms are not of interest here since they will not influence the inertia of a slow particle. Here $\vecr_{ab} =\vecr_b - \vecr_a$ is the vector from particle $a$ to particle $b$, and $r_{ab} = |\vecr_{ab}|$.

The equation of motion for particle $1$ is given by the Euler-Lagrange equation
\begin{equation}\label{eq.euler.lagrange}
\frac{\dfd}{\dfd t} \frac{\partial L}{\partial \vecv_1} = \frac{\partial L}{\partial \vecr_1} \; \Leftrightarrow\; \dot{\vecp}_1 = \vecF_1 .
\end{equation}
All terms involving accelerations will occur on the left hand side here, so this is what we need to calculate. Calculation gives,
\begin{equation}\label{eq.gen.momentum}
\frac{\partial L}{\partial \vecv_1} = m_1 \vecv_1 + \frac{G m_1}{2 c^2} \sum_{b=2}^N \frac{m_b}{r_{1b}} \left[ 6\vecv_1 - 7\vecv_b - \frac{\vecr_{1b} (\vecv_b \cdot \vecr_{1b}) }{r_{1b}^2}\right],
\end{equation}
for the, so called, generalized momentum $\vecp_1=\partial L/\partial \vecv_1$. Now assume that particle 1 is at the origin in a homogeneous isotropic expanding universe of density $\rho$, and with Hubble parameter $H$. The particles $m_b$ are then replaced by mass elements $\rho\, \dfd V$ of position $\vecr$ and velocity $\vecv = H \vecr + \vecu$. Here $\vecu$ is an overall velocity of the universe relative to the origin. We can then replace the sum in (\ref{eq.gen.momentum}) with an integral and get
\begin{equation}\label{eq.gen.momentum.int}
\vecp_1 = m_1 \vecv_1 + \frac{G m_1}{2 c^2} \int \frac{\rho}{r} \left[ 6\vecv_1 - 7(H \vecr + \vecu) - \frac{\vecr (H r^2 + \vecu\cdot\vecr) }{r^2}\right]\, \dfd V.
\end{equation}
We now calculate the integral on the right hand side.

We introduce spherical coordinates $(r, \varphi, \theta)$ and do the integration over the visible universe. At the radius $R$ of the visible universe the Hubble expansion leads to recession at the speed of light, $HR=c$. The volume element in spherical coordinates is $\dfd V = r^2 \sin\theta\, \dfd r\, \dfd\varphi\, \dfd\theta$. Without loss of generality we assume that $\vecu = u\vece_z$. Since $\vecr = r (\sin\theta\,\vece_{\varphi} + \cos\theta\,\vece_z)$, where $\vece_{\varphi}=\cos\varphi\,\vece_x + \sin\varphi\,\vece_y$, the scalar product term becomes
\begin{equation}\label{eq.calc.sclar}
\vecr (\vecu\cdot\vecr) = r^2 u (\sin\theta\,\vece_{\varphi} + \cos\theta\,\vece_z) \cos\theta.
\end{equation}
The integrations over the sphere of radius $R$ will make the terms involving $H$ vanish for symmetry reasons, since these are multiplied by $\vecr$. Nothing depends on the angle $\varphi$ in the integral so the term multiplying $\vece_{\varphi}$ also vanishes. Two different integrals then remain to calculate, first
\begin{equation}\label{eq.pot}
\int \frac{\rho\, \dfd V}{r} = 4\pi \rho \int_0^R  r \dfd r = 2\pi \rho R^2 ,
\end{equation}
and then
\begin{equation}\label{eq.int.scal}
\int \frac{\rho \cos^2\theta\, \dfd V}{r} = 2\pi \rho \int_0^R r\, \dfd r \int_0^{\pi} \cos^2\theta\, \sin\theta \,\dfd\theta = \frac{2}{3}\pi \rho R^2 ,
\end{equation}
due to the scalar product term.

This gives us the result
\begin{equation}\label{eq.gen.momentum.calculated}
\vecp_1 = m_1 \left[ \left(1+ 6 \frac{G\pi\rho R^2}{c^2} \right)\vecv_1 - \left( \frac{22}{3} \frac{G\pi\rho R^2}{c^2} \right) \vecu \right].
\end{equation}
In order to understand this, the meaning of the quantity $G\pi\rho R^2/c^2$, which we can rewrite,
\begin{equation}\label{eq.quatity.in.p}
\sigma \equiv \frac{G\pi\rho R^2}{c^2} = \frac{G \pi}{H^2}\rho ,
\end{equation}
using $R=c/H$, must be investigated.

Studying cosmology using general relativity and the assumption of an expanding homogeneous, isotropic universe one finds that there is a specific mass/ energy density $\rho_c$ that makes (three dimensional) space flat \cite{BKcheng_tp,jordan}. This density is called the critical density and is given by
\begin{equation}\label{eq.rho.crit}
\rho_c = \frac{3H^2}{8\pi G}.
\end{equation}
It is interesting to note that the critical density corresponds to the mass $M$ of the universe inside the Hubble radius $R=c/H$ being such that the Hubble radius is equal to the Schwarzschild radius $R=2GM/c^2$.

Comparing (\ref{eq.rho.crit}) with (\ref{eq.quatity.in.p}) we see that
\begin{equation}\label{eq.sigma.Omega}
\sigma = \frac{3}{8} \frac{\rho}{\rho_c} \equiv \frac{3}{8} \Omega ,
\end{equation}
where $\Omega$ is standard notation in cosmology for the ratio of the density to the critical density. The generalized momentum (\ref{eq.gen.momentum.calculated}) of particle 1 now becomes,
\begin{equation}\label{eq.gen.mom.omega}
 \vecp_1 = m_1 \left[ \left(1+ \frac{9}{4} \Omega \right)\vecv_1 - \left( \frac{11}{4} \Omega \right) \vecu \right].
\end{equation}
Returning to my formulation of Mach's principle in (\ref{eq.mach.princip.precise}) it is seen to be realized with this $\dot{\vecp}_1$ if $\Omega = 2$. For this value of the density ratio the generalized momentum is
\begin{equation}\label{eq.gen.mom.omega.2}
 \vecp_1 = m_1 \frac{11}{2} (\vecv_1 - \vecu ) \; \Rightarrow \; \dot{\vecp}_1 = m_1 \frac{11}{2} (\veca_1 - \vecg_{\rm u} ).
\end{equation}
{\it I.e.}\ for $\Omega=2$ the acceleration in Newton's second law is relative to the acceleration $\dot{\vecu}=\vecg_{\rm u}$ of the universe as a whole. One also notes that the "bare" mass $m_1$ has been "renormalized" to\footnote{If this renormalization is not permissible one can interpret Eq.\ (\protect\ref{eq.gen.mom.omega}) as requiring that all of the kinetic energy is due to interaction. Then the the $m_1 \vecv_1$-term in $\vecp_1$ vanishes and one concludes that the actual mass is $m=m_1 9\Omega /4$. Mach's principle would then require that the second term also has a factor $9/4$ instead of the $11/4$ obtained here.} $m=11 m_1/2$.

\section{Conclusions}
We see that while forces, arising from $\vecF_1 = \partial L/\partial \vecr_1$ in (\ref{eq.euler.lagrange}), decrease at least as $r^{-2}$, the inertial terms in $\vecp_1$ decrease only as $r^{-1}$. Consequently inertia has an intrinsic non-local nature. It is thus difficult to investigate by local measurements -- the main reason that these matters remain obscure and intimately connected to cosmology. This should be an important insight of these investigations.

In reference \cite{veto2} Vet\H{o} has found that the fictitious Coriolis force $-2m\,\vecomega\times\vecv^*$ can be understood as due to the gravitomagnetic field from the rest of the universe if $\Omega=1$. Support for such a standpoint comes also from investigations by Mart\'{i}n et al.\ \cite{martin&al} where numerical $\Omega$-values near 1 are found. Both these references, however, use linearized forms of general relativity that neglect the gravitational back-reaction of the accelerating particle on the background universe, whereas the EIH formalism used here retains back-reaction effects to linear order in the mass $m_1$. This makes the results of \cite{martin&al,veto2} comparatively unreliable.

In conclusion I have elucidated Mach's principle and found that a precise formulation of the principle can be consistent with the Einstein-Infeld-Hoffmann approximation to general relativity if the density of the universe is twice the critical density. This is in qualitative agreement with other investigations that find that Mach's principle requires the density parameter to be of order of magnitude unity. Already Berry (\cite{BKberry}, page 39) in his 1976 book found that simple estimates required that the density of the Universe should be $\rho_{\rm Berry} = H^2 /(2\pi G) = 4 \rho_c /3$ to obey Mach's principle. At that time the observed and inferred amount of galactic matter was only 4 percent of this value. Nowadays we definitely know that the order of magnitude of the mass/energy density of the Universe is such that Mach's principle is physically viable. \\ \\

\noindent{\bf\Large Acknowledgements}\\
I am grateful to Dr.\ Johan Sten for helpful comments. I also would like to thank the referee of European Journal of Physics for an unusually constructive, thorough, and clarifying report on a previous version of this manuscript.


\end{document}